# Trajectory Dynamics of Gas Molecules and Galaxy Formation

Pedro J. Llanos[1,*], James K. Miller[2] and Gerald R. Hintz[3]

1. Marie Curie Research Fellow, Tres Cantos, Madrid, Spain, 28760, Member AAS/AIAA
2. Retired Chief Scientist, Space Navigation and Flight Dynamics Section, KinetX Inc., Simi Valley, CA, 93065, Associate Fellow AIAA
3. Adjunct Professor, Astronautical Engineering Department, University of Southern California, Los Angeles, CA, 90089, Associate Fellow AIAA

**ABSTRACT**

The probability distribution of the velocity of gas molecules in a closed container is described by the kinetic theory of gases. When molecules collide or impact the walls of a container, they exchange energy and momentum in accordance with Newton's laws of motion. Between collisions, the trajectory of individual molecules is a straight line, neglecting gravity. During the formation of a galaxy, the stars are constrained to a region of space and exchange energy and momentum in a manner similar to molecules. In this paper, an exact model of an ideal gas is derived and analyzed to determine the probability distribution of the molecular velocities, which are then compared with the probability distribution of velocities associated with stars during galaxy formation.

**Key words**: gas: dynamics – gas: molecules – Boltzmann – Maxwell – galaxies: formation – galaxies: evolution – galaxies: ring – galaxies: spiral – galaxies: globular – galaxies: collision

## 1. INTRODUCTION

The velocity distribution of the molecules of a gas is described by the kinetic theory of gases. The trajectory of individual molecules is a straight line if gravity is neglected. When molecules collide or impact the walls of a container, they exchange energy and momentum in accordance with Newton's laws of motion. A statistical characterization of the velocity distribution was performed by Boltzmann and Maxwell during the end of the 19th century. There was much controversy during this time and Boltzmann proposed several theories to describe the probability distribution function (pdf) of the molecular velocities. The issues raised seemed to have been fermenting but were considered settled after Boltzmann's death in 1905. Unfortunately, these researchers did not have computers to test their theory and the purpose of this paper is to perform that test.

An exact model of the gas dynamics, while simple in principle, involves complicated programming to determine the pairs of molecules involved in collisions and the times of the collisions. Also, the computer time required to analyze a sufficiently large number of molecules and collisions is excessive, although not out of reach of today's high speed computers.

- E-mail: pedro.j.llanos.1@gmail.com

A simplified approximate model that analyzes only the velocity distribution would relate more readily to analytic theories, but is more difficult to justify. In this paper, three models are analyzed and compared with theory. The first yields a velocity magnitude distribution that matches the Maxwell distribution with high precision and is referred to as Model 1, the Maxwell model. The second is intended to improve the Maxwell model and yields a velocity distribution that is significantly different from the Maxwell distribution, but within the error of experimental verification and is referred to as Model 2, the velocity-dependent model. The third (Model 3) is an exact deterministic model that includes both position and velocity. The essential difference between the first two relates to the probability that two molecules will collide. The Maxwell model assumes all molecules are equally likely to collide at any given time and the velocity-dependent one assumes that the probability of collision is speed dependent.

### 1.1 Impact Velocity Model

The models that will be described here will all use the same impact velocity model illustrated in Figure 1. The velocities of molecule 1 and 2 before impact are $\mathbf{V}_1$ and $\mathbf{V}_2$, respectively, and the velocities after impact are $\mathbf{U}_1$ and $\mathbf{U}_2$ respectively, as illustrated on the left side of Figure 1.

If $\mathbf{\Delta V}_1$ and $\mathbf{\Delta V}_2$ are the changes in velocity, we have:

$$\mathbf{U}_1 = \mathbf{V}_1 + \mathbf{\Delta V}_1 \qquad (1)$$
$$\mathbf{U}_2 = \mathbf{V}_2 - \mathbf{\Delta V}_2 \qquad (2)$$

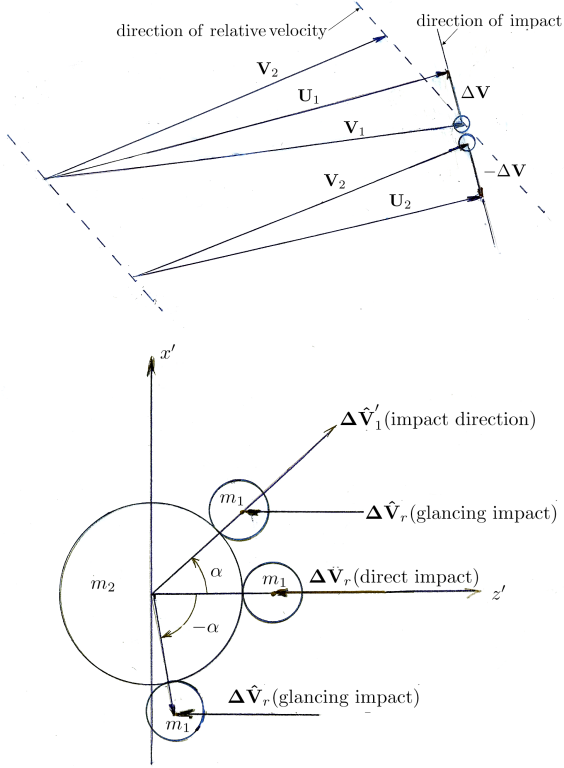

**Figure 1.** Molecule collision geometry.

and

$$|\Delta \mathbf{V_1}| = \frac{2(\mathbf{V_2}\cdot \Delta \hat{\mathbf{V}}_2) - 2(\mathbf{V_1}\cdot \Delta \hat{\mathbf{V}}_1)}{1 + \frac{m_1}{m_2}} \quad (3)$$

$$|\Delta \mathbf{V_2}| = \frac{m_1}{m_2}|\Delta \mathbf{V_1}| \quad (4)$$

The impact velocity direction is illustrated at the bottom of Figure 1. The $\hat{\mathbf{z}}'$ axis is in the direction of the relative velocity vector. The $\hat{\mathbf{x}}'$ axis is in the reference $\hat{\mathbf{x}} - \hat{\mathbf{z}}$ plane. The $\hat{\mathbf{y}}'$ axis completes the right-hand Cartesian coordinate system. The scattering angles $\alpha$ and $\beta$ define the impact direction. When $\alpha$ is zero, the impact is direct, and when $\alpha$ is not zero, the impact is glancing. The range of $\alpha$ is from 0 to $\pi$. The angle $\beta$ defines the rotation of the impact direction around the $\mathbf{z}'$ axis and has a range from 0 to $2\pi$. In the primed coordinate system, the impact direction is defined by the unit vector,

$$\Delta \hat{\mathbf{V}}_1' = \begin{bmatrix} \sin\alpha\cos\beta \\ \sin\alpha\sin\beta \\ \cos\alpha \end{bmatrix} \quad (5)$$

and the relative velocity vector is

$$\Delta \mathbf{V_r} = \mathbf{V_1} - \mathbf{V_2} \quad (6)$$

The velocity relative coordinate system defined above has the following coordinate axes unit vectors:

$$\hat{\mathbf{x}}' = \frac{\Delta \hat{\mathbf{V}}_r \times \hat{\mathbf{z}}}{|\Delta \hat{\mathbf{V}}_r \times \hat{\mathbf{z}}|}$$

$$\hat{\mathbf{y}}' = \frac{\Delta \hat{\mathbf{V}}_r \times (\Delta \hat{\mathbf{V}}_r \times \hat{\mathbf{z}})}{|\Delta \hat{\mathbf{V}}_r \times (\Delta \hat{\mathbf{V}}_r \times \hat{\mathbf{z}})|} \quad (7)$$

$$\hat{\mathbf{z}}' = \Delta \hat{\mathbf{V}}_r$$

The 3x3 transformation matrix from the velocity vector fixed to the reference coordinate system is given by

$$T = [\hat{\mathbf{x}}' \, \hat{\mathbf{y}}' \, \hat{\mathbf{z}}'] \quad (8)$$

The unit vector in the impact direction in the reference coordinate system is given by

$$\Delta \hat{\mathbf{V}}_1 = T \, \Delta \hat{\mathbf{V}}_1' \quad (9)$$

and

$$\Delta \mathbf{V_1} = |\Delta \mathbf{V_1}| \, \Delta \hat{\mathbf{V}}_1 \quad (10)$$

## 2. COLLISION MODELS

### 2.1. Maxwell and Velocity-Dependent Models

The collision models described here will all operate on a database containing the state of each molecule. This data base is a table where the row dimension corresponds to the molecule number and the columns contain the molecule state. The state columns include velocity, position, radius, mass and other molecule-dependent parameters that may be of interest. The first model makes use of the fact that collisions occur one at a time. If the pair of molecules involved in the next collision can be identified, then it is not necessary to keep track of position. Model 1 assumes all molecules are equally likely to participate in the next collision. These participants are selected by a random number generator. Once the participants have been identified, it is necessary to determine the scattering angles $\alpha$ and $\beta$. Since the molecules are assumed to be small relative to the distance between them, the point at which the relative velocity vector pierces the plane perpendicular to the relative velocity vector and passes through the center of the molecule may be assumed to be uniformly distributed. The scattering angles are thus given by the following equations:

$$\sin\alpha = \sqrt{Ran(0,1)}$$
$$\beta = 2\pi \, Ran(0,1) \quad (11)$$

where $Ran(0, 1)$ is a random number uniformly distributed between 0 and 1.

Model 1 will yield a probability distribution that is Maxwellian to very high precision. The problem with Model 1 is that the molecules are not equally likely to collide as has been assumed. Model 2, the velocity-dependent model, makes the assump-tion that the selection of molecule participants is velocity-dependent. During the time between collisions, all the molecules lay down a track whose length is equal to the velocity magnitude multiplied by the time. The probability that a collision will occur is proportional to the length of these tracks. The first participant is selected at random weighted by the track length or velocity magnitude. The second participant is at the end of the track and is selected at random as for Model 1. This simple change to Model 1 yielded a pdf that is different from Maxwellian by a few percent as will be shown later. This result was surprising and begs the question as to which model is correct.

## 2.2. Deterministic Model

The third model, referred to as the deterministic model, was implemented for the purpose of verifying the statistical assumptions of the first two models. The third model computes explicitly both the position and velocity of each molecule and thus eliminates the need for random number generators and assumptions about the positions of molecules. The algorithm is complex and involves many computations for each collision. First, Model 3 is initialized with the position, velocity, mass and radius of each molecule. Two columns are added to the position and velocity array that control molecule selection. The first column (NTARG) contains the number of the molecule that is in line to be impacted or minus one for a wall. Some time in the future every molecule must impact another molecule or a wall. The second column (TIMPACT) contains the time to impact. The position-dependent collision geometry is illustrated in Figure 2. The relative position and velocity for each pair of molecules are given by,

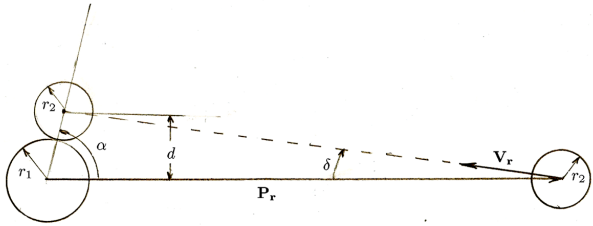

**Figure 2.** Position-dependent molecule collision geometry.

$$\mathbf{P_r} = \mathbf{P_1} - \mathbf{P_2}$$
$$\mathbf{V_r} = \mathbf{V_1} - \mathbf{V_2} \quad (12)$$

The angle ($\delta$) between the relative velocity vector and the negative of the position vector is given by

$$\cos\delta = -\frac{\mathbf{P_r} \cdot \mathbf{V_r}}{P_r V_r} \quad (13)$$

Collision occurs if the following inequality is satisfied:

$$P_r \sin\delta < r_1 + r_2 \quad (14)$$

If collision occurs, the scattering angle, $\alpha$, is approximately

$$\sin\alpha = \frac{P_r \sin\delta}{r_1 + r_2} \quad (15)$$

The exact $\alpha$ may be obtained by solving the following two equations iteratively using the above approximation for $\alpha$ as an initial guess. The distance between two colliding particle flight paths is

$$d = \frac{\tan\alpha \tan\delta \, P_r}{\tan\alpha + \tan\delta} \quad (16)$$
$$\sin\alpha = \frac{d}{r_1 + r_2}$$

and the time to collision is

$$\Delta t = \frac{d}{\sin\delta \, V_r} \quad (17)$$

The scattering angle $\beta$ around the relative velocity vector is given by

$$\beta = \arctan\left[\frac{(\mathbf{V_r} \times \mathbf{P_r}) \cdot \hat{\mathbf{y}}'}{(\mathbf{V_r} \times \mathbf{P_r}) \cdot \hat{\mathbf{x}}'}\right] - \frac{\pi}{2} \quad (18)$$

where $\hat{\mathbf{x}}'$ and $\hat{\mathbf{y}}'$ are the unit vectors defining the x' and y' axes of the velocity relative coordinate system. The vector normal to the relative position and velocity vectors leads the scattering direction by 90 degrees. The scattering direction is in the plane of the relative position and velocity vectors.

A container is defined in the shape of a rectangular parallelepiped. For each molecule (i) the time of impact is computed for each wall of the parallelepiped. This time is simply the distance to the wall divided by the x, y, or z component of velocity. The minimum time is entered in TIMPACT(i) and a minus one is placed in NTARG(i). For each i,j pair of molecules the time to impact is computed as described above. If this time is less than TIMPACT(i), then TIMPACT(i) is updated with the new time and NTARG(i) is set equal to j, the number of the current molecule. Thus, n(n-1) determinations must be made for the first collision, where n is the total number of molecules. This computation will take several seconds of computer time for the first collision. The TIMPACT array is then searched for the time of the next collision. The positions of all of the molecules are then propagated to the time of the next collision. If the collision is with a wall, the velocity vector is reflected off the wall. If the collision is with another molecule, the angles $\alpha$ and $\beta$ are computed as described above, the differential velocity resulting from impact is computed, and the velocities of the two participants are updated. The NTARG array is searched and NTARG(i) is set equal to zero for the participants in the collision and any other molecule that had a participant as a target. The above algorithm is repeated for the next collision. Those molecules with NTARG(i) equal to zero must be searched for new targets, greatly reducing the number of calculations needed.

The deterministic model is an exact model for an ideal gas, the so-called billiard ball model. However, billiard balls are not exactly ideal. If we fill a gymnasium with billiard balls in outer space, they will collide but eventually lose their kinetic energy after several collisions. They will also acquire some spin or angular rotation energy. Molecules in a container will not lose their kinetic energy as a result of collisions with each other because of quantum energy states. Molecules do not warm up or get dents. Also, since most of the mass is concentrated in the nucleus, molecules do not acquire much spin energy. The electrons are much less massive than protons or neutrons. Molecules thus probably behave more like an ideal gas than billiard balls. However, there is one important difference. Molecules do not collide like billiard balls or an ideal gas. They are subject to, what is called an R5 potential. The scattering angle α is thus a function of mass and relative velocity. Simulations of various potential models reveal that the scattering angle has no affect on the final steady state velocity distribution. Symmetry and the central limit theorem produce a near normal pdf for velocity components and a near Maxwellian pdf for velocity magnitude.

## 3. MONTE CARLO RESULTS

A Monte Carlo simulation of 200,000 molecules in a container was simulated for 50,000,000 impacts. On the average, each molecule is engaged in 100 impacts. The initial velocity distribution is arbitrary because the steady state solution is of interest. In order to achieve steady state with a minimum number of impacts, each of the three velocity components was initialized with a normal distribution. The root mean square (rms) velocity was set equal to one. Figure 3 shows the resulting velocity magnitude distribution for the Maxwell model and the velocity-dependent model.

The top histogram is the result for the Maxwell model. The abscissa is velocity magnitude scaled to 50 histogram bins. Thus bin 50 would correspond to a velocity of 3.108, the maximum velocity obtained. The ordinate exhibits the probability of a molecule being in a particular bin times the total number of molecules or simply the number of molecules in each bin.

The integral from zero to infinity is thus 200,000, the sum of all the histogram bins. Also plotted is the exact Maxwell distribution function scaled the same way. For the Maxwell model, the fit is so good that these curves are on top of one another and cannot be distinguished. Model 1 yields an "exact" Maxwell distribution.

The histogram at the bottom of Figure 3 is the result for Model 2, the velocity-dependent model. The heavy line is the velocity-dependent model distribution histogram and the fine line is the Maxwell distribution function. The agreement is not very good, indicating that either Model 2 or the Maxwell distribution function is in error.

The deterministic model, Model 3, was executed and produced the same results as for Model 2, thus providing an important verification of the Model 2 assumptions.

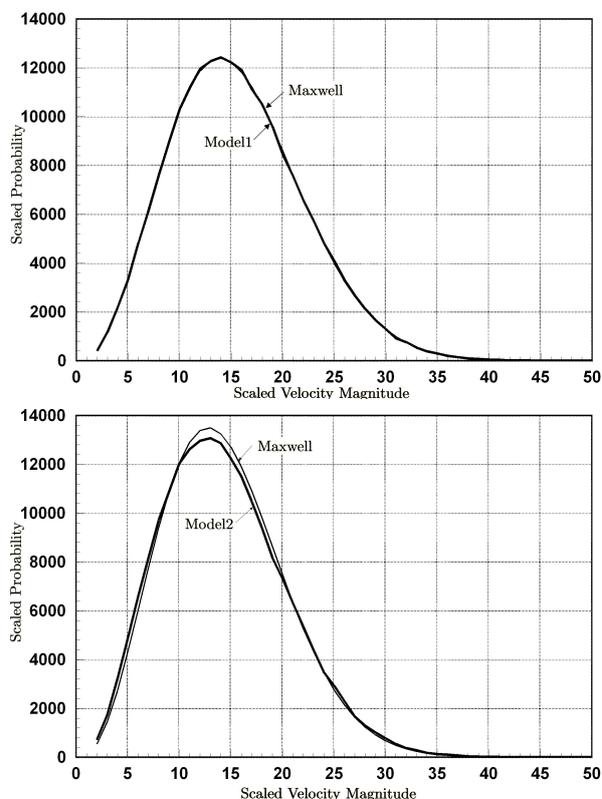

**Figure 3.** Maxwell model and velocity dependent model histograms

Since the differences between the Model 2 and Model 3 histograms and Maxwell distribution function shown in Figure 3 are too small to be seen clearly, they are plotted on the bottom of Figure 4. The Maxwell distribution is not the fundamental distribution function of interest. The top of Figure 4 shows the differences between the z component of velocity histogram and the normal probability distribution function. The histogram bins span the z velocity component from about -2.1 to +2.1 and zero corresponds to bin 25.

Since there is no preferred direction, a true signature in the data must be symmetrical about zero velocity. The Model 2 and Model 3 results show a significant signature that is symmetrical about zero, indicating that the distribution function is not normal. Also shown in Figure 4 is the result of evaluating an empirical formula, which will be discussed later.

Since the Model 1 results for a velocity component have a normal pdf and the coefficients of the binomial theorem, or Pascal's triangle, also have a normal pdf, we may obtain some insight into the dynamics of the collisions by identification with Pascal's triangle.

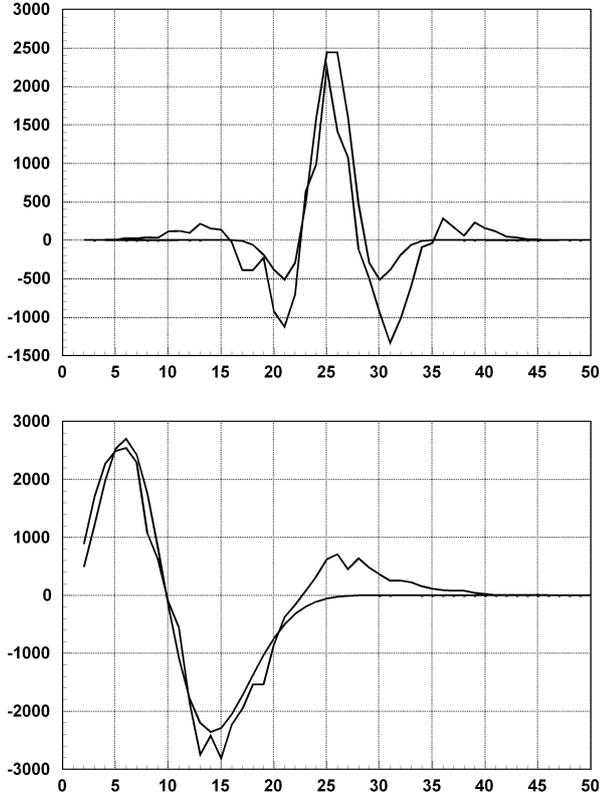

**Figure 4.** Model differences from Maxwell and normal distribution functions.

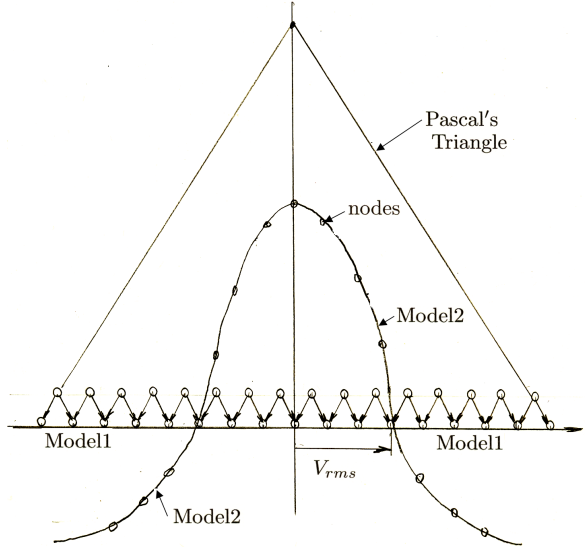

**Figure 5.** Modified Pascal's triangle.

If the Model 1 results can be reduced to a coin flipping strategy, then the Model 2 results may be explained as a variation of this strategy. This strategy is illustrated in Figure 5.

A problem with the Pascal triangle simplified model is that molecules are not equally likely to collide. The faster molecules are involved in more collisions. Thus, more faster molecules make it to the next generation than the slower molecules. This fact is illustrated in Figure 5 by the Model 2 set of nodes. The Model 2 pdf corresponds to the values of the nodes on this curve and is not normal. The shape of the Model 2 curve is constrained to be symmetrical because there is no preferred direction and it must have the same integral from minus infinity to plus infinity as the Model 1 curve. The curve shown at the top side of Figure 4 appears to be a damped cosine function so, as a first guess, we may modify the normal pdf as follows:

$$\frac{\delta P(V_z)}{\delta V_z} = \sqrt{\frac{m}{2\pi KT}} \left[ \exp\left(\frac{-mV_z^2}{2KT}\right) + \sum_{i=1}^{2} D_i \cos\left(C_1 \frac{V_z}{\sqrt{\frac{2KT}{m}}}\right) \exp\left(\frac{-C_2 mV_z^2}{2KT}\right) \right] \quad (19)$$

Extending the one-dimensional pdf to three dimensions, the probability of being in a box of dimension $\delta V_x$ $\delta V_y$ $\delta V_z$ is given by

$$\delta P(V) = \frac{\partial P(V_x)}{\partial V_x} \frac{\partial P(V_y)}{\partial V_y} \frac{\partial P(V_z)}{\partial V_z} \delta V \quad (20)$$

where

$$\delta V = \delta V_x \delta V_y \delta V_z \quad (21)$$

Because of spherical symmetry, the probability (P(V)) of being in a box defined by the volume element $\delta V$ is the same for a given V. Here, the region of velocity space defined by arranging all the velocity vectors about a common origin is euphemistically referred to as a volume element. Thus, all the volume elements in a shell with a given velocity magnitude will have the same probability. The volume of a shell is given by $4\pi V^2 \delta V$ and we may arbitrarily select the volume element defined by $\delta V^2 = 3\delta V_z^2$, where all three components of velocity are equal. The pdf for V is thus

$$\frac{\delta P(V_z)}{\delta V_z} = 4\pi V^2 \left(\frac{m}{2\pi KT}\right)^{3/2} \left[ \exp\left(\frac{-mV^2}{6KT}\right) + \sum_{i=1}^{2} D_i \cos\left(C_1 \frac{V}{\sqrt{\frac{6KT}{m}}}\right) \exp\left(\frac{-C_2 mV^2}{6KT}\right) \right]^3 \quad (22)$$

The coefficients $D_1 = 0.03141$, $D_2 = 0.02859$, $C_1 = 6.202$ and $C_2 = 8.76$ are all non-dimensional and scaled to apply for all temperatures and gas molecular weights. Observe that if the coefficients $D_1$ and $D_2$ are set equal to zero, the Maxwell distribution is obtained,

$$\frac{\delta P(V)}{\delta V} = 4\pi V^2 \left(\frac{m}{2\pi KT}\right)^{3/2} \exp\left(-\frac{mV^2}{2KT}\right) \quad (23)$$

## 4. EXPERIMENTAL RESULTS

An experimental verification of Maxwell's probability distribution function is documented in Reference 1. The experiment consisted of heating Potassium and Thallium in an oven to 900 Celsius and venting the atoms through a velocity selector into a detector. The velocity selector was a cylinder with a slot on one side and a curved slot on the other that rotated at a variable rate of up to 4000 rpm such that atoms would enter the curved slot and cross the cylinder at various transit times, depending on their velocity and the location of the detector along the axis of the cylinder. The detector would measure the intensity, or number of atoms, at the velocity corresponding to position along the axis of the cylinder or the angle that the cylinder would rotate dependent on the rotation rate. The device was quite sophisticated, being cooled with liquid nitrogen and sealed to provide a high vacuum.

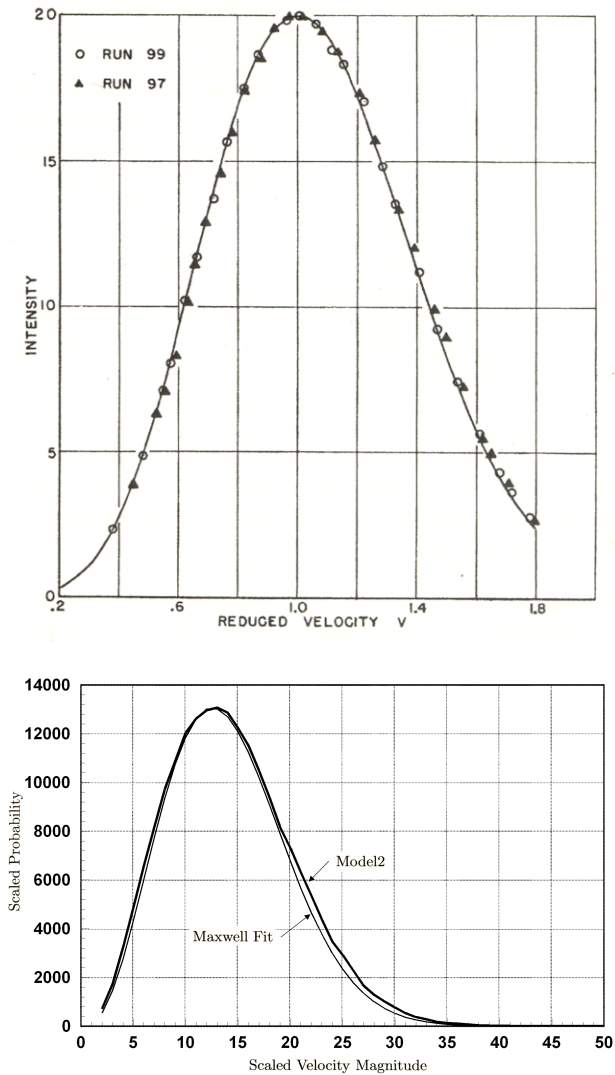

**Figure 6**. Experimental results and theory comparison.

The results of the Thallium experiments are shown at the top of Figure 6. Also plotted at the bottom of Figure 6 is the theoretical Maxwell pdf. The agreement with the Model 2 theory developed in this paper does not, at first, appear very good. Inspection of this comparison shown on Figure 3 reveals a gap between Maxwell's theory and the Monte Carlo results at the peak intensity or most probable velocity. This gap, if it exists, would not show up in the experimental results because the scaling between intensity and actual number of molecules is not known precisely. Therefore, the experimental results were scaled to force them to equal the Maxwell theory at the peak. Also, the temperature was adjusted so that the low velocity experimental results matched Maxwell's theory. The problem with these adjustments is that the Maxwell pdf is a single parameter theory. Once the temperature is fixed, the probabilities and shape of the curve are fixed.

If the gap shown in Figure 3 is real, the area between the curves at the peak will be distributed elsewhere. This redistribution is evident at the top of Figure 6. Close inspection of the high velocity side of the curve reveals that the experimental results are above the Maxwell theory prediction. If these results are accurate, there is a problem, because the integral of the Maxwell pdf from zero to plus infinity is one and the experimental results are greater than one. R. C. Miller and P. Kusch acknowledged this problem in Reference 1. The authors say, "It is seen that the largest discrepancies occur on the high velocity side of the maximum, where there is a small excess of atoms in the experimental distribution. It should be noted that the experimental points could be plotted with the high velocity side matched to the theoretical curve. The intensities at the maximum velocity would no longer coincide and the experimental distribution would then appear to be deficient of atoms on the low velocity side." In other words, the theoretical curve could be shifted to the left by adjusting the temperature, but the area under the curve would still be greater than one. Another modification that would give a better fit and correspond to the Model 3 simulation would be to raise the theoretical curve at the peak a small amount.

Another approach that is relatively simple to implement would be to scale the Maxwell distribution to force the peak to coincide with the Model 3 simulations and adjust the temperature to match the curves on the low velocity side. These results could then be compared directly with the experimental results. The bottom of Figure 6 shows the results of this procedure. The experimental results at the top of Figure 6 depart a small amount from the theoretical Maxwell distribution at a reduced velocity of 1.4. The Model 3 results on the bottom part of Figure 6 also depart a small amount. At a reduced velocity of 1.8, the experimental results trend back to the Maxwell curve. The Monte Carlo results also trend back. The obvious conclusion is that the experimental results confirm the Model 3 simulation. However, since these results are within the measurement error and are obscured by the relatively small sample size, a more accurate experiment is needed. However, it should be noted that the experimental results may be expected to underperform the theoretical results because of leakage from the apparatus. Since they overperform when compared with the Maxwell distribution, the detector must be counting atoms that are not there. Since particle detectors are presumed to measure with high precision, the Model 3 simulation seems more consistent with intuition. The experiment documented in Reference 1 was performed in 1955.

## 5. GALAXY FORMATION

### 5.1. Spiral and Ring Galaxies

An ideal gas in a container has a different pdf than a gas not in a container, for example, gas expanding into a vacuum. For a gas not in a container, the faster molecules move farther away from the origin, a greater distance than the slower molecules. The faster molecules are proportionally farther apart than the slower molecules and thus collide less frequently. The result is that slow molecules and fast molecules are more or less equally likely to collide and the pdf tends toward Maxwellian. For numerical simulations conducted on a sample of about 250,000 molecules, the pdf tended toward Maxwellian, but the collisions stopped after all the molecules were on trajectories that moved radially outward and, therefore, did not intersect.

The Model 3 equations assume rigid walls that surround the gas. The Earth's atmosphere has the same distribution as Model 3 where the walls are defined by the Earth's surface and layers of the atmosphere of constant density are constrained by gravity. Stars in a galaxy are also constrained by gravity and it would be interesting to compare the pdf of star velocities with the pdf of molecule velocities. The equations of motion for stars in a galaxy and molecules in a container share some striking similarities. First, they both are constrained by conservation of energy and momentum. Second, the dynamics of collisions and the dynamics of gravitational encounter result in the bodies exchanging energy and momentum and thus being deflected the same way. The attractive force of gravity will swing one body around the other from one side to the other and the repulsive force of collision will deflect a body from the same side as the impact. There are some important differences in the dynamics that will have no effect on the pdf. First, the stars never collide. They are so small relative to the distances between them that the probability of collision is virtually zero. Entire galaxies can collide without individual stars colliding. Second, the gravitational attraction of molecules is insignificant. However, there is one significant difference between colliding molecules in a container and gravitationally attractive stars in a galaxy. Conservation of angular momentum results in the stars retaining the angular momentum they are initialized with as they form into a galaxy. Molecules in a container surrender their net momentum in any direction to the walls of the container. The net angular momentum in any direction becomes zero. The momentum is absorbed by the container and, for the Earth, results in a small perturbation to the spin axis and rotation rate.

A simple modification of Model 3 will enable simulation of both molecules in a container and stars in a galaxy. We used Newton's third law to obtain the force between particles as in Eq. 24. The force exerted on a particle $i$ is computed as the sum of the forces from the rest of the particles $j \neq i$ in the system. Although this approach provides the correct acceleration of every particle it takes of the order of $N^2$ calculations, which is very time consuming. For simplicity we will assume that all the stars have equal mass normalized to one. The force exerted on one single particle by the N-1 other particles is

$$\overline{F}_{ij} = -GM_i \sum_{j \neq i}^{N-1} \frac{M_j}{r_{ij}^2} \overline{r}_{ij} \qquad (24)$$

where the position vector between two particles is:

$$\overline{r}_{ij} = (x_i - x_j)\hat{x} + (y_i - y_j)\hat{y} \qquad (25)$$

In the N-body problem, the particles move according to Newton's equations of motion. The mass of each star is assumed to be the same and equal to one solar mass. These are arbitrary parameters subject to change depending on the galaxy model. The initial cluster of stars can also be chosen arbitrarily for a certain ellipticity value. In this paper, we analyze several different galaxy examples for different ellipticities, $\varepsilon$, and number of stars. The position and velocity of each star is obtained by integrating the acceleration,

$$\begin{aligned} a_i^{(n)} &= \frac{\overline{F}_i^{(n)}}{m_i} \\ v_i^{(n+1)} &= v_i^{(n)} + a_i^{(n)} \Delta t \\ x_i^{(n+1)} &= x_i^{(n)} + v_i^{(n+1)} \Delta t \end{aligned} \qquad (26)$$

Every particle must be initialized with a position and velocity in the galaxy. The initial conditions are expressed in polar coordinates $(r,\theta)$ that may be converted to $(x,y)$ coordinates:

$$\begin{aligned} x &= r \cos\theta \\ y &= \varepsilon \, r \sin\theta \end{aligned} \qquad (27)$$

The angle $\theta$ is randomly determined between 0 and $2\pi$ and the initial r-coordinate is also randomly distributed between 0 and ¼. The initial velocities $(v_x, v_y)$ of the particles can be determined by taking the derivative of the positions:

$$\begin{aligned} v_x &= V_r \cos\theta - RV_\theta \sin\theta \\ v_y &= V_r \sin\theta + RV_\theta \cos\theta \end{aligned} \qquad (28)$$

Non-dimensional units are assumed for the simulations. $V_r$ is assumed to be zero and the tangential velocity is assumed to be 50 scale velocity units. The integration time step is fixed at $10^{-4}$ and the gravitational constant is assumed to be 0.05. For galaxy formation, the inverse square gravitational attraction between each body is included, the walls are moved far away and the radii of the bodies are made vanishingly small. For molecules in a container, the gravitational attraction is set to zero or turned off. Computing the initial condition for molecules in a container is not a problem. Any initial velocity distribution will quickly evolve into the same steady state pdf for any gas. Even the extreme assumption that one molecule has all the energy and all the other molecules are at absolute zero temperature will achieve the same steady state. For a cluster of stars, the initial assumption of angular momentum and star density will have a profound effect on the outcome. Figure 7 shows some typical results of galaxy formations. We start from a very simple assumption of position and velocity distribution.

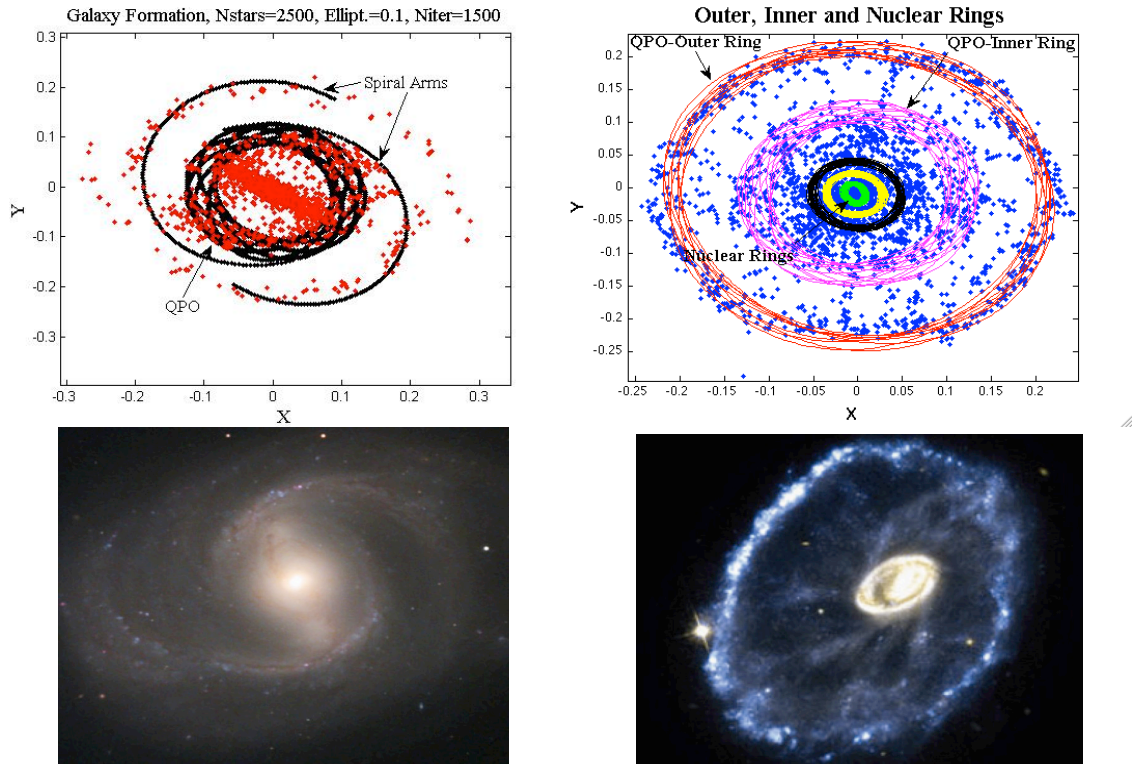

**Figure 7.** Simulated galaxy formation and actual galaxies. The spiral galaxy shows spiral arms and a quasi-periodic orbit after 1500 integration steps. The real spiral galaxy is M91 (SB). The ring galaxy simulation, after 1500 integration steps, illustrates near periodic orbits around the outer and inner rings and several nuclear rings coexisting together very close to the center of the ring galaxy giving evidence of nuclear activity at the center of the galaxy. The real ring galaxy is the cartwheel galaxy.

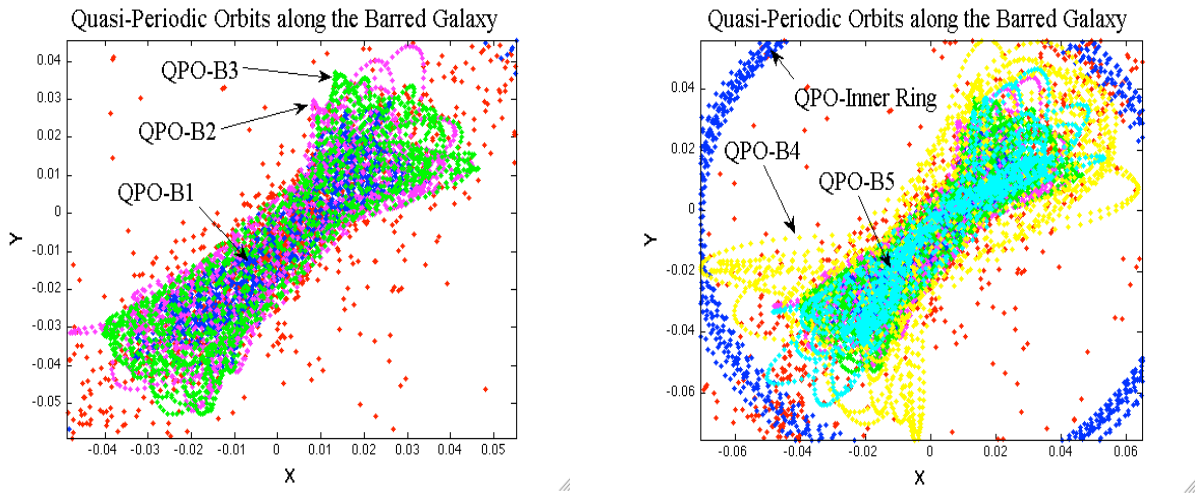

**Figure 8.** Quasi-Periodic Orbits: QPO-B1 (blue), QPO-B2 (magenta), QPO-B3 (green), QPO-B4 (yellow), QPO-B5 (cyan) in a barred galaxy. Red dots are the simulated stars forming the bar of the galaxy. The spiral galaxy simulation was obtained after 500 integration steps.

Assuming the stars formed from an expanding universe, individual clusters may have the positions of the stars uniformly distributed in spherical shells and slowly rotating about the center of mass at a fixed angular rate. The left side of Figure 7 shows a spiral galaxy simulation (Llanos (2) and Llanos (7)) with 2500 stars with an ellipticity of the initial bulge of 0.1. After 1500 integration steps, we obtained the corresponding spiral arms, which mimic the simulated data very closely. In this case, the spiral arms winding around the galaxy are longer and closer to the center of the galaxy due to the lower density of stars that they carry.

This simulation mimics the real galaxy formation, M91 shown on the lower left side of Figure 7, with spiral arms moving counterclockwise and a clear bar.

Top left corner of Figure 7 shows a quasi-periodic orbit (QPO) or near periodic orbit that supports the bar inside corotation. This orbit forms a ring at the same time that it supports the spiral structure outside corotation. These orbits (Heggie and Hut (11)) can give us important information to better understand the role of resonances in the support of spiral patterns in galaxies, but the study of resonances will not be covered in this paper. Also shown on the right side of Figure 7 is an actual photograph of a ring galaxy and the results of galaxy simulation.

Figure 8 shows quasi periodic orbits for a barred galaxy. In an early stage of our galaxy simulation (500 integration steps), the length of the bar extends to the inner ring. We can find chaotic orbits around these rings. These orbits adopt a dog-bone shape, that is, they look like straight bars in the middle but they expand sideways towards the end of the bar before reaching the quasi-periodic orbit formed around the inner ring. Therefore, due to the fact that there is a larger conglomeration of stars on the center of the bar, highly elongated orbits are formed about the center of the galaxy. We can see this exchange of material (Contopoulos (4)) in the form of rapid oscillations or quasi-periodic orbits (Figure 8). The fact that there is a bar formed is an indication that the old stellar disk of the galaxy is heavy (Contopoulos (4), Block, Puerari, Freeman, Groess and & Block (9)). In our computer simulations, we found a family of quasi-periodic orbits that extend along the bar of the galaxy, that is, these orbits are lined up with the long axis of the bar (Sparke and Gallagher (10)). Unlike the stars from the spiral arms that behave like density waves, most of the stars remain within the bar itself. While the motion of the stars in the spiral follows a near-circular motion, the stars in the bar no longer follow near-circular motions but other chaotic orbits (peanut shape or dog-bone shape) as we can see in Figure 8. In the left side of Figure 8, we show a clear example of three orbits, QPO-B1 (blue), QPO-B2 (magenta) and QPO-B3 (green) aligned along the bar. For the same galaxy formation (right side of Figure 8), we explored other orbits QPO-B4 (yellow) and QPO-B5 (cyan). QPO-B5 is the most elongated and narrow orbit along the bar axis extending to the size of the diameter of the QPO-Inner ring. QPO-B4 is also an orbit formed in the bar but it tends to expand outwardly on the sides of the bar connecting with the QPO Inner ring.

In our second simulation, we used an original formation of 3500 stars with an ellipticity factor of 1.0 and obtained the galaxy depicted in the upper right-hand side of Figure 7 after 1250 integration steps. This galaxy simulation shows a clear outer ring (red), an inner ring (magenta) and the bulge of the galaxy. In this simulation, rings of different sizes coexist together. The outer and inner rings depicted in Figure 7 have nearly circular shapes and the nuclear rings are slightly oval shaped. Both outer and inner rings are connected through many long 'spurs' or filaments of different orientations, which are thought to be the channels through which material may be transported between the inner and outer rings. This transport behaves like a pulse providing orbital support of stars living in that ring. Both nuclear and non-nuclear rings were obtained over 2000 integration steps.

For actual galaxies, we cannot obtain precise measures of the velocity of individual stars from the photographs. We cannot even see all of them. For the simulation, we know the precise number of stars and their positions and velocities at all times. A histogram of velocity magnitudes resembles a Maxwell distribution. In Figure 9, the difference between the best fit Maxwell distribution and the simulated pdf is plotted for a spiral galaxy, a ring galaxy and a globular galaxy or a star cluster.

Also shown in Figure 9 is the pdf difference for molecules in a container. The abscissa is normalized velocity magnitude and the ordinate is called mapped entropy. Since the Maxwell distribution represents maximum disorder, mapped entropy, which is the deviation from the Maxwell distribution, represents order. The more mapped entropy shown in Figure 9 the more order. The entropy of every star or molecule shown in Figure 9 can be uniquely computed from the value of the mapped entropy. Therefore, the mapped entropy maps one-to-one on the true entropy as defined by Boltzmann. Thus, the ring galaxy and the spiral galaxy show more mapped entropy or order consistent with the photograph and simulation.

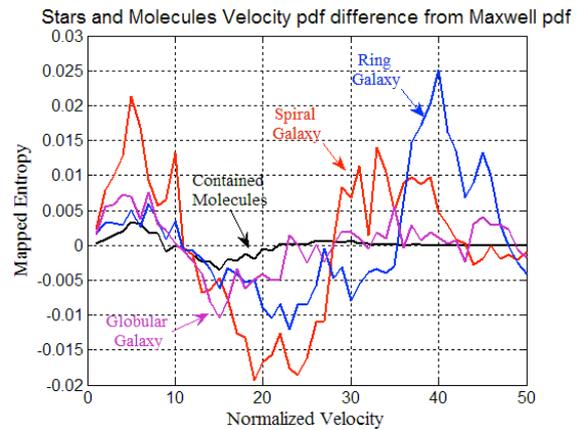

**Figure 9.** Star and molecule velocity pdf difference from Maxwell pdf.

Molecules in a container show very little mapped entropy. What is striking about Figure 9 is the signature. Galaxies and molecules share the same basic signature only different amplitudes.

*5.2. Globular Galaxy*

The globular or cluster galaxy has a pdf that is closest to molecules in a container. Since the angular momentum is small, these galaxies exhibit little structure or order. However, the pdf of a globular galaxy can only approach the pdf of molecules in a container because of the residual angular momentum that cannot be disposed of by interaction with a container. Figure 10 shows the simulation results after formation of a globular galaxy of 5000 stars after 1500 integration steps.

The top of Figure 11 shows a plot of the evolution of the pdf from an initial simple uniform distribution to a quasi-steady state distribution. Also shown is the best fit Maxwell distribution.

The fit is good on the high side but exhibits a characteristic signature on the low-velocity side. There are more stars with velocity magnitudes below the average rms velocity and a deficiency of stars above the rms velocity. This is the same result as was obtained by molecules in a container and the reason is the same. Faster moving stars or molecules engage in more collisions or encounters than slower moving ones. The faster moving bodies are more likely to loose energy and thus there is an excess of slow bodies relative to the Maxwell distribution.

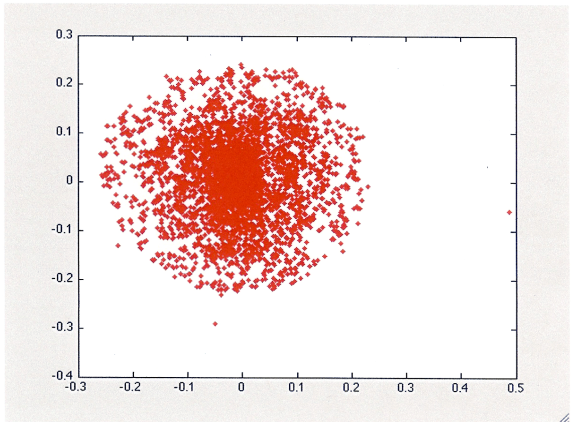

**Figure 10.** Simulated Globular Galaxy of 5000 stars after 1500 integration steps.

One may question whether the final steady state solution shown at the top of Figure 11 is really steady state or the pdf slowly evolving to some other distribution. The bottom part of Figure 11 shows the galaxy evolution for an initial condition that is forced to be Maxwellian. The velocity magnitudes of the final pdf shown at the top of Figure 11 were modified to be Maxwellian.

The procedure for modifying the velocity magnitudes was designed to minimize the change in angular momentum. A Maxwellian distribution was computed with the same rms velocity as the final distribution. Both sets of velocity magnitudes were ordered from the smallest to the greatest. The Maxwell distribution was transferred in the same order to the final pdf shown on the bottom of Figure 11 and provides the initial pdf. The simulation evolved back to the original pdf thus demonstrating that the final solution is near steady state.

The difference between the Maxwell distribution and the final steady state solution for the globular cluster galaxy and molecules in a container is shown in Figure 12. The similarity of the two signatures cannot be attributed to statistical coincidence. Furthermore, the results for the molecules and galaxy formation were completed by the authors of this paper using completely independent software. The work was completed before the author's were aware of each other's existence. The comparison shown here could only be made in non-dimensional units.

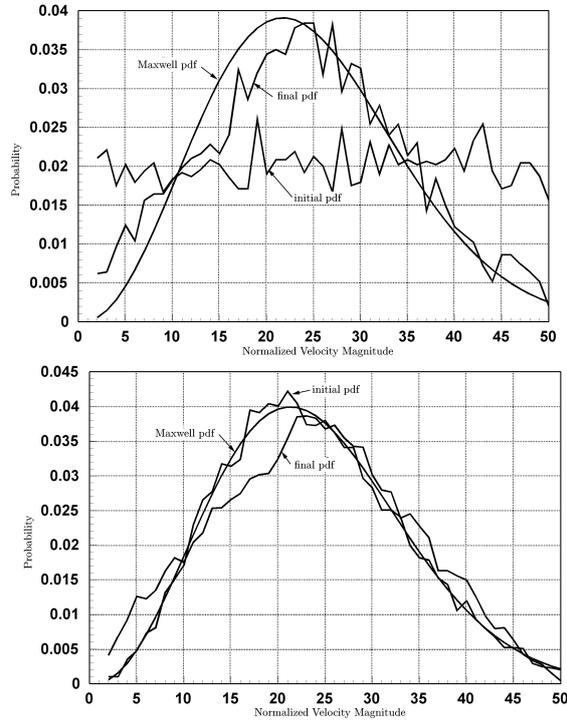

**Figure 11.** Evolution of pdf for Simulated Globular Galaxy Formation.

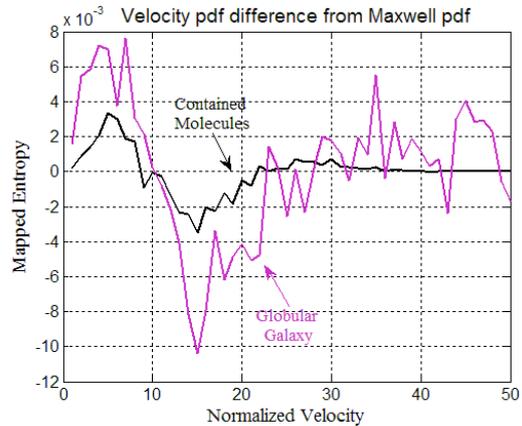

**Figure 12.** Comparison of Molecule and Globular Galaxy Velocity Distributions.

A galaxy is typically several hundred thousand light years in diameter and evolves over a time span of about 50 million years. The molecules in a container were assumed to be at standard temperature and pressure and thus separated by about 30 diameters. The width of the walls of the container was several microns or about the width of a human hair. For any arbitrary initial velocity distribution, steady state is achieved after about 20 collisions. The time to reach steady state is measured in micro seconds. After normalization and plotting at the same scale, the molecules and galaxies have similar signatures.

Similar to the three-body problem, the N-Body gravitational problem (Woolfson and Pert (5), Baron (6), Lambers & Lof (8), Heggie & Hut (11)), such as the evolution of a stellar cluster, was analyzed showing very chaotic behaviors. Many of the global features of galaxies such as bars, rings, and arms may be explained by the chaotic transport and resulting morphology controlled by the invariant manifolds (Athanassoula, Romero-Gómez & Masdemont (3)) of unstable periodic orbits in the N-Body Problem.

Figure 13 displays the evolution of a spiral galaxy (Llanos (2) & Llanos (7)) with 2500 stars initially distributed with an ellipticity of 0.1. The vector field is intended to give the reader a better idea about the direction of the stars during the evolution of the galaxy. Frame 0 is the initial distribution of stars; frame 40 shows how the distribution of stars has slightly rotated counter-clockwise; and frame 125 shows conspicuous spiral arms emanating from the ends of the bar. These spiral arms continue evolving until 1500 integration steps (Figure 13). Early stages of the galaxy formation show that the stars inside the corotating zone bend inwards as shown by the velocity vector fields and stars outside the corotating zone have velocity vectors pointing outwardly. As the galaxy evolves after 250 integration steps, the stars' (in the arms) velocity vectors are more aligned to the near circular motion of the spiral arms depicted in frames with 875 and 1500 integration steps in Figures 13.

There are two main scenarios (Block, Puerari, Freeman, Groess & Block (9)) that form polar rings: mergers of two galaxies or external gas accretion. In our simulations, we illustrate a merger of two spiral galaxies forming a polar ring galaxy. The two merging spiral galaxies represent a cluster of 4500 stars and result in the formation of a ring galaxy as illustrated by frame 3000 in Figure 14 after several collision encounters. The red elliptical galaxy has initially 2500 stars distributed with a 0.1 ellipticity whereas the blue elliptical galaxy has 2000 stars initially distributed with a 0.3 ellipticity. The red galaxy is centered at (0,0) and the blue galaxy has an offset of (0.25,-0.25). It is interesting to see in frame 2500 a broad stream of stars emerging at the left of the cluster. These escapers reach a maximum distance along the negative x-axis, then they are slowly channeled back (see frame 2750) to the center of the galaxy. Finally, most of these escapers are near the center of the ring galaxy as illustrated in frame 3000.

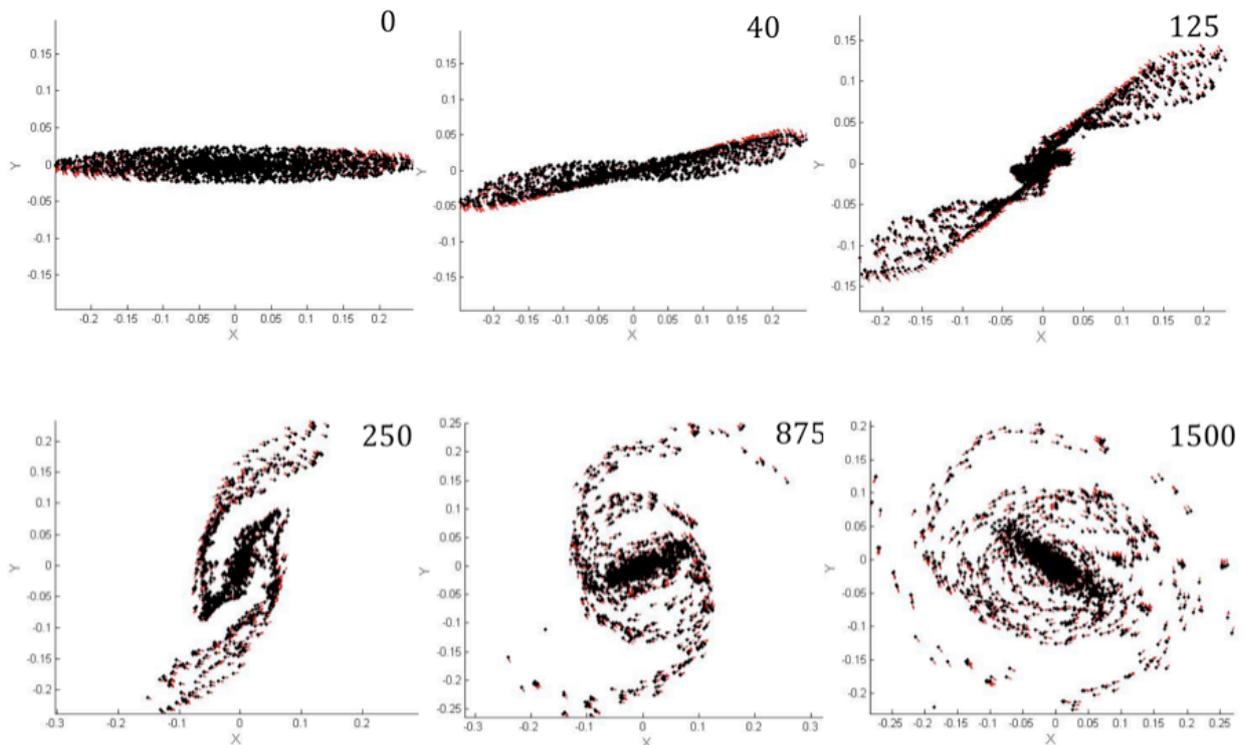

**Figure 13.** Vector field (red arrows) of spiral galaxy formation are illustrated at different time integration steps as indicated in the top right corner of each frame. We indicate frames 0, 40, 125, 250, 875 and 1500. In this galaxy, we can see the formation of spiral arms and a small bulge of stars at 125 integration steps. This bulge of stars evolves into a bar with more defined spiral arms at 250 integration steps. Then, a clear bar forms at 875 integration steps along with a ring. This ring evolves into outer and inner rings at 1500 integration steps with a more elongated bar and a better-structured set of spiral arms.

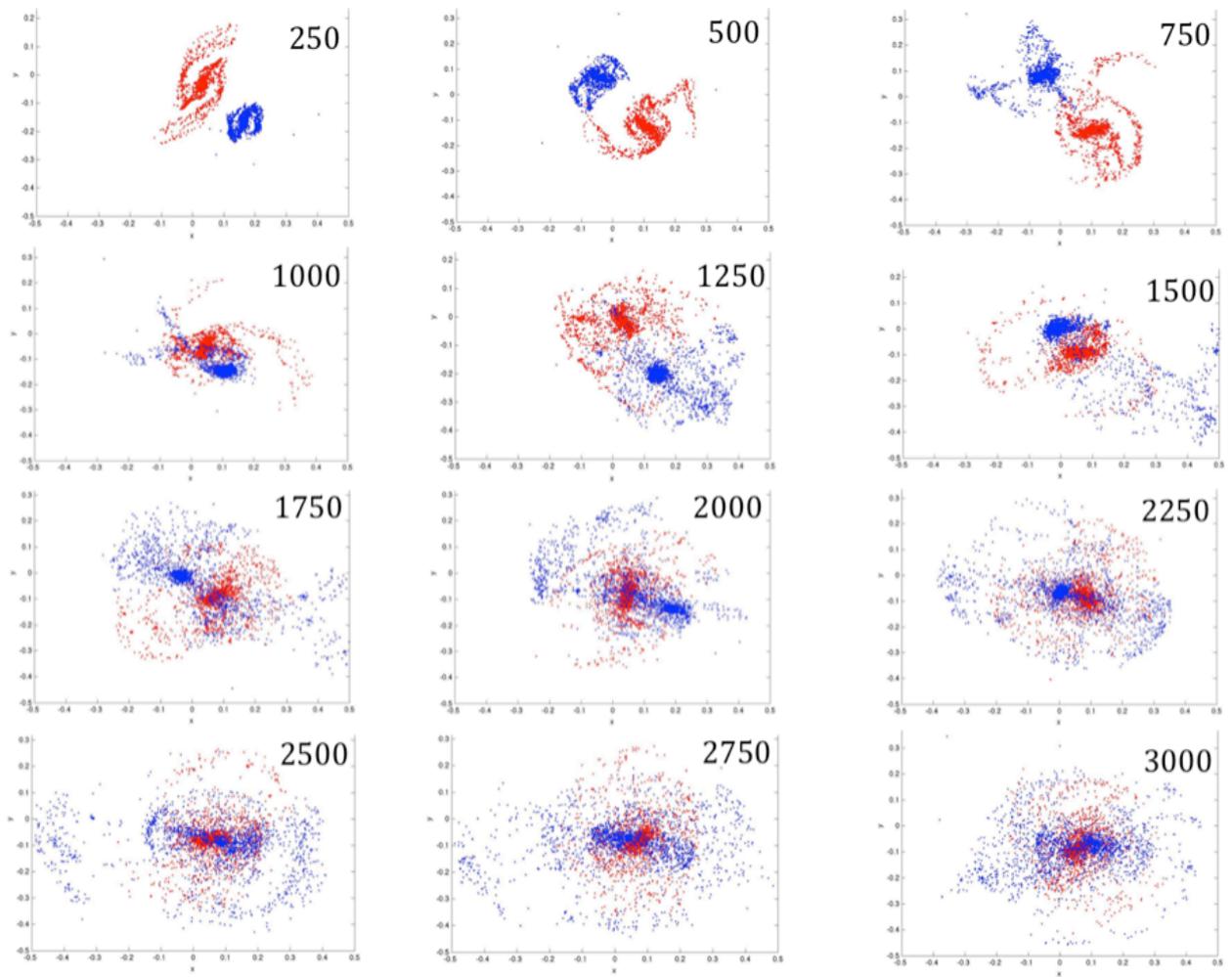

**Figure 14.** Galaxy evolution of two spiral galaxies displayed at different time integrations. The galaxy evolution is indicated from left to right and top to bottom so that the frame of the galaxy in the top left corner occurs at 250 integration steps and the galaxy in the bottom right corner evolved after 3000 integration steps.

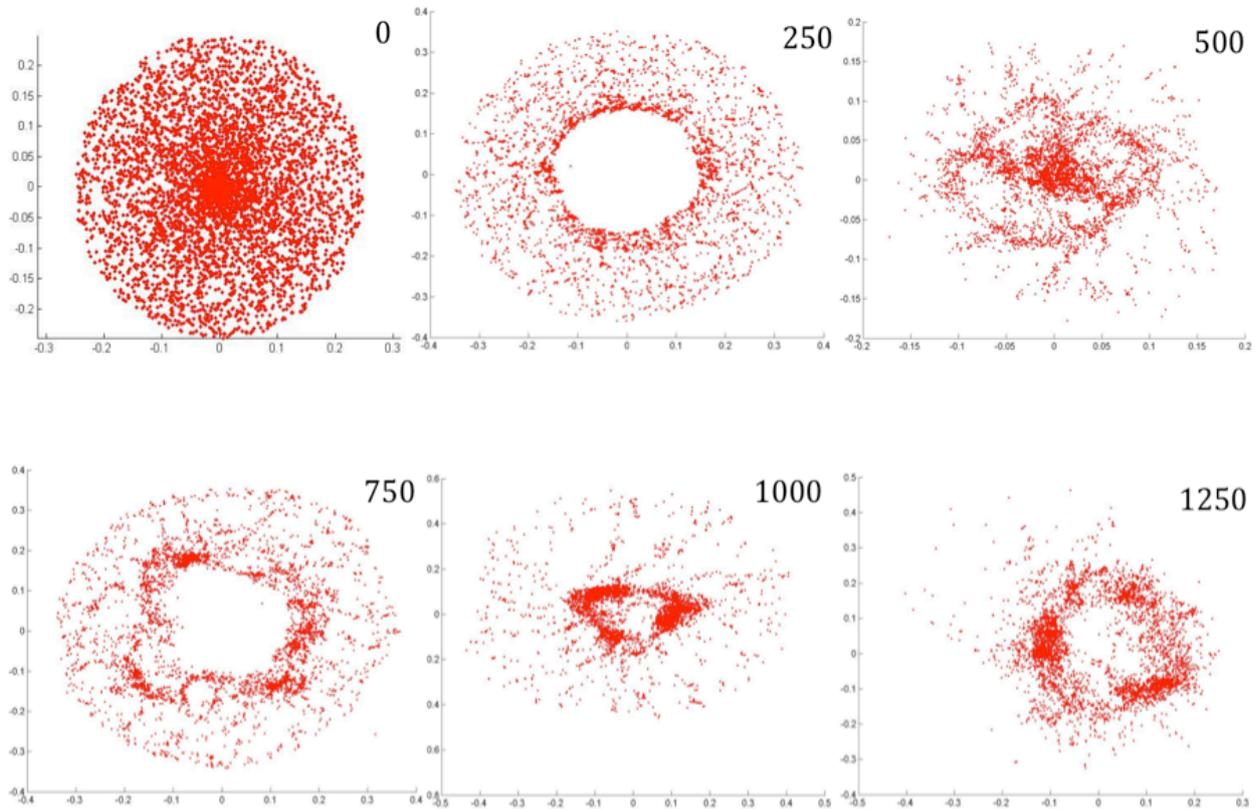

**Figure 15.** Ring hollow galaxy of 5000 stars initially distributed with an ellipticity of one. In this simulation, we are assuming a non-radial velocity of 10 non-dimensional units and the same tangential velocity used in the rest of the simulations (50 non-dimensional units).

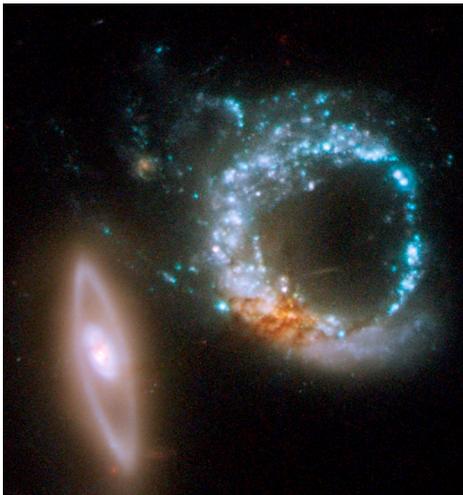

**Figure 16.** Real optical image of the Arp 147 Ring Galaxy obtained by the Hubble Space Telescope in 2008. This galaxy was produced when a spiral galaxy (right) collided with an elliptical galaxy (left).

Finally, we show a hollow ring galaxy depicted in Figure 15. The initial distribution of 5000 stars has an ellipticity factor of one. The stars were given an initial velocity of 50 non-dimensional units and 10 non-dimensional units in the tangential and radial directions, respectively. We show frames 0, 250, 500, 750, 1250 and 1500 in Figure 15. Frame 250 shows a clear formation of a ring galaxy of stars. This cluster of stars pulsates inwards forming a ring and spurs in frame 500. The cluster of stars pulsates outwards forming another ring. This process continues repeating itself and, in frame 1500, the cluster of stars adopts a similar form to that of the real Arp 147 ring galaxy.

The enhanced blue colors of the remnant ring galaxy in Figure 16 indicate new forming massive stars after the collision occurred. Some of these stars ended up in supernova explosions leaving behind either black holes or neutron stars. The lower left part of the blue galaxy shows a reddish area that is thought to be the location where the galaxy was originally hit when the elliptical galaxy passed through the spiral galaxy on the right.

## 6. SUMMARY AND CONCLUSIONS

A Monte Carlo analysis of the kinetic theory of gasses has revealed a discrepancy between Maxwell- Boltzmann theory and results obtained by statistical analysis. These results have been supported by theory that identifies the random impact changes in velocity with the binomial coefficients and thus the normal probability distribution. The conclusion of this paper is that the velocity distribution of components of velocity in an ideal gas is not a normal Gaussian contrary to the theory of Boltzmann. This discrepancy carries over to the Maxwell distribution function. Experimental measurements of velocity distribution conducted in 1955 by R.C. Miller and P. Kusch support this conclusion even though the measurement accuracy is not statistically strong enough to be conclusive.

In this paper, we also explored and examined the morphology and dynamics of barred, ring and globular galaxies as well as collision between galaxies using a simple Direct N-Body Newtonian Model. In our simulations, we showed that within these global structures we could find orbits where transport of material takes place. Even though this method is a simple approach, it gives us very interesting features that could explain some of the unsolved problems in galactic dynamics and globular clusters.

From a qualitative point of view, we were able to have a better understanding on how some of these galaxy features, such as the formation of the spiral arms, the bar, and the rings, arose. Moreover, we saw that there exist connections in the form of spurs and chaotic orbits as a way of material transport between the rings.

Our study shows very similar signatures when comparing the molecules and galaxies pdf distribution indicating a slight difference from the Maxwellian pdf distribution.

All the results obtained in this paper assumed Newtonian motion. General relativity, gas, dark matter and dark energy are assumed to be small or insignificant.

**REFERENCES**
1. Miller, R. C. and Kusch, P.,Velocity Distributions in Potassium and Thallium Atomic Beams, Physical Review, Vol 99. No 4. August 15, 1955.
2. Llanos, P., J., Morphology and Dynamics of Galaxies, AIAA Region VI Student Conference, San Diego State University, CA, March 24-26, 2011 - Honorable Mention.
3. Athanassoula, E., Romero-Gomez, M., Masdemont, J. (2009) Rings and Spirals in Barred Galaxies - I. Building Blocks, MNRAS, 394, pp. 67-81.
4. Contopoulos G. (2002) Order and Chaos in Dynamical Astronomy, Springer-Verlag, Berlin Heidelberg, Germany.
5. M. M. Woolfson and G. J. Pert (1999) An Introduction to Computer Simulation, Oxford University Press, New York.
6. Edward Baron, Numerical Methods (unpublished notes), Fall 2003, University of Oklahoma.
7. Llanos, P. (2011) Morphology and Galaxy Dynamics, Third Annual GPSS Poster Symposium, University of Southern California.
8. Lambers, J., Lof, H. (2008) Assignment 3, Assignment N-body: The gravitational Nbody Problem, CME212 Introduction to Large-Scale Computing in Engineering, Institute for Computational & Mathematical Engineering, Stanford University, http://stanford.edu/class/cme212/.
9. David L. Block, Ivanio Puerari, Kenneth C. Freeman, Robert Groess and Elizabeth K. Block (2004) Penetrating Bars Through Masks of Cosmic Dust, Astrophysics and Space Science Library, Springer, Dordrecht, The Netherlands.
10. Linda S. Sparke and John S. Gallagher (2000) Galaxies in the Universe, Cambridge University Press, Cambdrige, United Kingdom.
11. Douglas Heggie and Piet Hut (2003) The Gravitational Million-Body Problem, Cambridge University Press, Cambridge, United Kingdom.